# Circularly polarized light scattering imaging of a cancerous layer creeping under a healthy layer for the diagnosis of early-stage cervical cancer

Nozomi Nishizawa[1*], Mahiro Ishikawa[1], Mike Raj Maskey[1], Asato Esumi[1], Toshihide Matsumoto[2], Takahiro Kuchimaru[3]

[1]*Department of Physics, School of Science, Kitasato University, Kanagawa 252-0373, Japan.*

[2]*Department of Medical Laboratory Sciences, School of Allied Health Sciences, Kitasato University, Kanagawa 252-0373, Japan.*

[3]*Division of Bioconvergence, Center for Molecular Medicine, Jichi Medical University, Tochigi 329-0498, Japan.*

*E-mail: nishizawa.nozomi@kitasato-u.ac.jp

**ABSTRACT**

**Significance:** Cervical cancer progresses through cervical intraepithelial neoplasia (CIN), which are precursor lesions of cervical cancer. In low-grade CIN, atypical cells generate inside the squamous epithelium, which causes the accuracy of cytodiagnosis for cervical cancer not to be very high. The grade of CIN can be estimated by the depth of atypical cell infiltration from the basal layer to the surface, rather than the abnormality of cells. Therefore, a non-invasive method is required to evaluate the depths of abnormal cells hidden at depths.

**Aim:** Cancerous tissues beneath healthy tissues were experimentally identified by using circularly polarized light scattering (CiPLS). This method enabled the changes in the size of the cell nuclei within the penetration depth in tissue to be investigated.

**Approach:** Artificial unexposed cancerous tissues were prepared that consisted of healthy/cancerous/healthy layers with various thicknesses of the topmost healthy layer and the cancerous layer. A polarization imaging camera with a quarter-wave plate was used to create distribution images of the circular polarization of the scattered light.

**Results:** CiPLS images indicated that the thickness variation of the top healthy layer (the depth of the cancerous layer) caused significant changes in the degree of circular polarization.

**Conclusions:** The depth of unexposed cancer lying within the optical penetration depth can be evaluated using a circular polarization imaging system based on the CiPLS method. These findings will lead to the development of a non-invasive optical diagnostic method for early-stage cervical cancer, potentially improving early detection and treatment outcomes.

## I. Introduction

Cervical cancer has both high morbidity and mortality; more than 600,000 women are diagnosed and more than 300,000 die every year worldwide [1, 2], making cervical cancer the fourth most frequent cancer among women globally [3]. Approximately 95 % or more of cervical cancers are caused by sustained infection with human papillomavirus (HPV) in the uterine cervix. These infections progress through cervical intraepithelial neoplasia (CIN) or adenocarcinoma *in situ* (AIN), precursor lesions of cervical cancer, to invasive cervical cancer (ICC) [4]. The uterine cervix consists of two distinct epithelia: the columnar epithelium of the endocervix and the squamous epithelium of the ectocervix, which are joined at the squamocolumnar junction (SCJ) (Figure 1 (a)). In the squamous epithelium near the SCJ, where cell proliferation is rapid, sustained HPV infection causes the generation of atypical cells, known as squamous intraepithelial dysplasia, at the bottom of the cervical squamous epithelium immediately above the basal layer separating it from the stroma (Figure 1 (c)). This condition is referred to as CIN. CIN is classified into three stages based on the ratio of abnormal cells in the cervical lining (Figure 1 (b)) [5]. When abnormal cells occupy more than one-third of the epithelial lining, mild CIN (CIN1) (Figure 1 (d)) shifts to moderate CIN (CIN2) (Figure 1 (e)). Abnormal cells occupying more than two-thirds of the epithelial lining corresponds to severe CIN (CIN3) (Figure 1 (f)). When the spread of abnormal cells reaches the surface of the epithelial layer, known as carcinoma *in situ*, its direction of spread turns inward, progressing to invasive cancer. The evolution of malignant cervical lesions is not unidirectional; the progression or regression between the grades of CIN occurs bidirectionally because of host immune protection. However, CIN, AIN, and early-stage cervical cancer are not associated with subjective symptoms. Therefore, even if the CIN grade is 1 or 2, periodic inspections are required to monitor the condition.

The initial diagnosis of cervical cancer involves a cytodiagnostic examination, in which the cervix is gently scraped with a spatula or brush to confirm the presence of atypical squamous cells of undetermined significance (ASC-US) [6, 7]. If the ASC-US are HPV-positive, the examination proceeds to a highly sensitive histopathological inspection using a colposcopy. However, for low-grade CIN, the cells collected during cytodiagnosis may not be from the area where the abnormal cell spread is the most advanced, as these cells are not exposed on the surface. Therefore, the accuracy of cytodiagnosis is not very high for low-grade CIN, but it is sufficiently high for high-grade CIN. As mentioned above, the classification of CIN grades is based on the ratio of atypical cells in the cervical squamous epithelium, determined by their depth rather than cell abnormality. Consequently, a non-invasive method is required to evaluate the depth of abnormal cells hidden at depths of a few submillimeters within the cervical squamous epithelium (Figure 1 (c)).

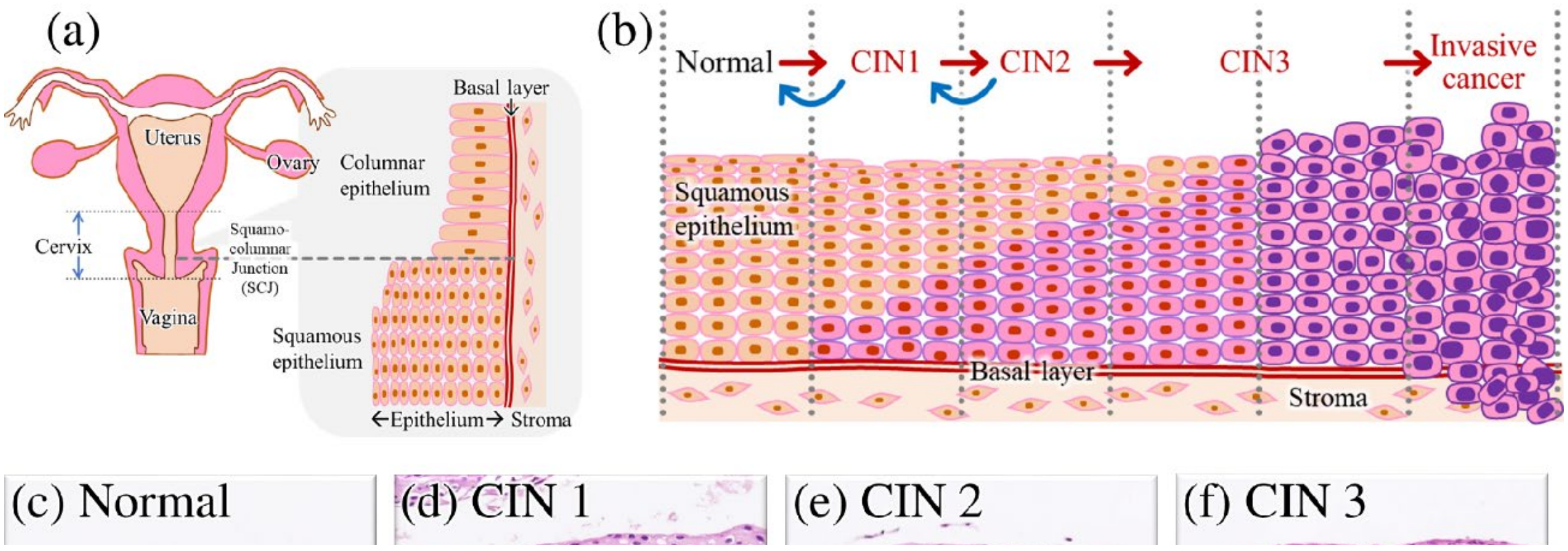

Figure 1: Schematic illustrations of the (a) cervix and uterus with the tissue structure near the SCJ and (b) cervical pre-cancer and cancer progression: normal, CIN1, CIN2, CIN3, and invasive cancer. Representative hematoxylin and eosin images of the squamous epithelium in (c) normal (d) CIN1, (e) CIN2, and (f) CIN3 cases.

Circularly polarized light scattering (CiPLS) can be used to investigate changes in the size of particles in turbid media [8, 9]. In the Mie scattering regime, where the scatterers are larger than the incident wavelength, the depolarization of polarized light due to multiple scattering in turbid media strongly depends on the ratio of the wavelength to the diameter of the particles. Thus, the degree of polarization of the light scattered from the media indicates the variation in the size of the scatterers [10]. Moreover, circularly polarized light (CPL) has greater persistence than linearly polarized light (LPL) in multiple Mie scattering [11]. Since forward scattering is dominant in the Mie scattering regime, linear polarization of light is drastically reduced by multiple scattering events that cause the rotation of the polarization plane, whereas circular polarization is not much affected [11, 12]. Therefore, CPL can retrieve information about scatterers from deeper. By applying the CiPLS method in biological observation, changes in the size of cell nuclei in biological tissues can be detected [8, 9, 13−15]. In most cancerous or precancerous tissues, cell nuclei that are approximately twice the size of normal cell nuclei can be found. Our previous study showed that CPL beams with wavelengths near 600 nm cause stronger depolarization of normal cell nuclei (smaller scatterers) compared to cancerous ones (larger scatterers). In contrast, at approximately 900 nm, cancerous cell nuclei show significant depolarization. [8, 9, 15]. These near-infrared light beams can penetrate to a depth of approximately 3 mm without complete depolarization [8, 15]. The depth at which LPL can be emitted from the surface without complete depolarization is comparatively shallow, whereas CPL which reaches the deeper layer and then goes back to the surface can retain the polarization. Therefore, when the scattering volume includes cancerous cells partially, the scattered light of CPL can indicate the ratio of enlarged cell nuclei in abnormal cells in the optical scattering volume of some mm in depth [16, 17]. The CiPLS method is characterized by being non-invasive, non-staining, and *in situ*, offering both surface and spatial resolution. It has the potential to evaluate squamous intraepithelial dysplasia non-invasively without the need for staining and any biomarkers.

In our previous study, computational analyses using the Monte Carlo simulation method for the CPL scattering process were performed for cancerous and healthy tissues, as well as bilayers consisting of them [15]. Calculations for a buried cancerous layer beneath a healthy layer show that the degree of circular polarization (DOCP) exhibits behavior depending on the depth of the cancerous layer. These calculations assume that optical parameters are independent to the tissue condition and cancer stage. Under these assumptions, the estimated measurable cancer depth is approximately 1.6 mm. This detectable depth is sufficient to diagnose the squamous epithelium whose typical thickness is approximately 0.7 mm.

In this study, we aimed to explore the potential of the CiPLS method for evaluating abnormal cells hidden at depths of a few millimeters. By experimentally demonstrating the identification of cancerous tissues beneath healthy tissues using an imaging system capable of circular polarization imaging, we provide a foundational step towards the optical detection of squamous intraepithelial dysplasia. This research could pave the way for more accurate and non-invasive diagnostic techniques in the future.

## II. Methods

Biological tissue samples were prepared by embedding sliced healthy and cancerous tissues in agarose using a VF-510-0Z vibrating microtome (Precisionary Instruments, NC, USA). Cancerous tissues were obtained from tumors harvested from a murine xenograft model established by the subcutaneous injection of human pancreatic cancer SUIT2 cells.

Table1: Sample list

| (mm) | 1 | 2 | 3 | 4 | 5 | 6 |
|---|---|---|---|---|---|---|
| $T_1$ | 0.0 | 0.5 | | | 1.0 | 1.5 |
| $T_2$ | 1.0 | 0.1 | 0.5 | 1.0 | 1.0 | 1.0 |
| $T$ | > 3 | | | | | |

Healthy tissues were obtained from the hind limb muscles of immunodeficient SCID mice used in the xenograft models. All animal experiments were approved by the Animal Experiment Committee of Jichi Medical University and were carried out in accordance with the relevant national and international guidelines. The muscle tissues are fibrous. The orientation of the fibrous tissues affects the azimuth angle of the depolarized light, however, the DOCP values exclude these anisotropic contributions of the tissues.

Figure 2 (a) shows the structure of the biological tissue. The thicknesses of the uppermost healthy tissue layer, buried cancerous tissue, and total tissue were $T_1$, $T_2$, and $T$, respectively. The total thickness $T$ of each sample was greater than 3 mm, a thickness through which the infrared light used in this study can hardly transmitted. The sample thicknesses are listed in Table 1.

Polarized light can be decomposed into two components that oscillate coherently and perpendicularly. The polarization state of light in transverse electromagnetic waves is typically described by a four-element vector known as the Stokes vector $\boldsymbol{S}$, $\boldsymbol{S} = (S_0, S_1, S_2, S_3)^T$, where $S_0$, $S_1$, $S_2$, and $S_3$ are the Stokes polarization parameters [18]. The first Stokes parameter, $S_0$, is the total intensity of the light; the second parameter, $S_1$, quantifies the preponderance of horizontal LPL over vertical LPL to the reference plane; the third parameter, $S_2$, provides the preponderance of $+45°$ LPL over $-45°$ LPL to the reference plane; and the last parameter, $S_3$, quantifies the preponderance of right-handed CPL over left-handed CPL. The DOCP value typically used is $|S_3|/S_0$; however, it is $S_3/S_0$ in the CiPLS method because the polarity of the scattered CPL, the sign of $S_3$, is also meaningful.

Figure 2 (b) shows the optical setup used for circular polarization imaging. The unpolarized and incoherent light from LEDs with wavelengths $\lambda$ of 617 nm and 850 nm and with 1.0 W and 1.6 W (M617L5 and M850LP1; Thorlabs, Inc.) was converted to right-handed CPL through a linear polarizer, half-wave plate, and quarter-wave plate (QWP) for the relevant wavelength. These CPL beams were irradiated onto the sample placed 10 cm away from the LED with incident angles of $\pm 30°$. When the angle was smaller than $30°$, contributions of a specular reflection from the surface increase significantly, causing the detected DOCP values to close to negative $(-1)$. In contrast, with a larger incident angle, light beams found it difficult to reach the deep layer. Additionally, the initial DOCP values of the incident CPL decreased because the incident efficiencies of the s-wave and p-wave components of CPL became different. When CPL was irradiated with $30°$, the spot shapes were ellipsoidal and extended in the direction of the incident angle. The minor and major axes of the elliptical spot were 25 mm and 29 mm, respectively. The DOCP value at the center of the spots was controlled to be $+1.00$. At the farthest circumference from the light source, the DOCP value was $+0.880$ at $\lambda = 617$ nm and $+0.955$ at $\lambda = 850$ nm, and the stability of the DOCP in the spot was 88.0% and 95.5%, respectively.

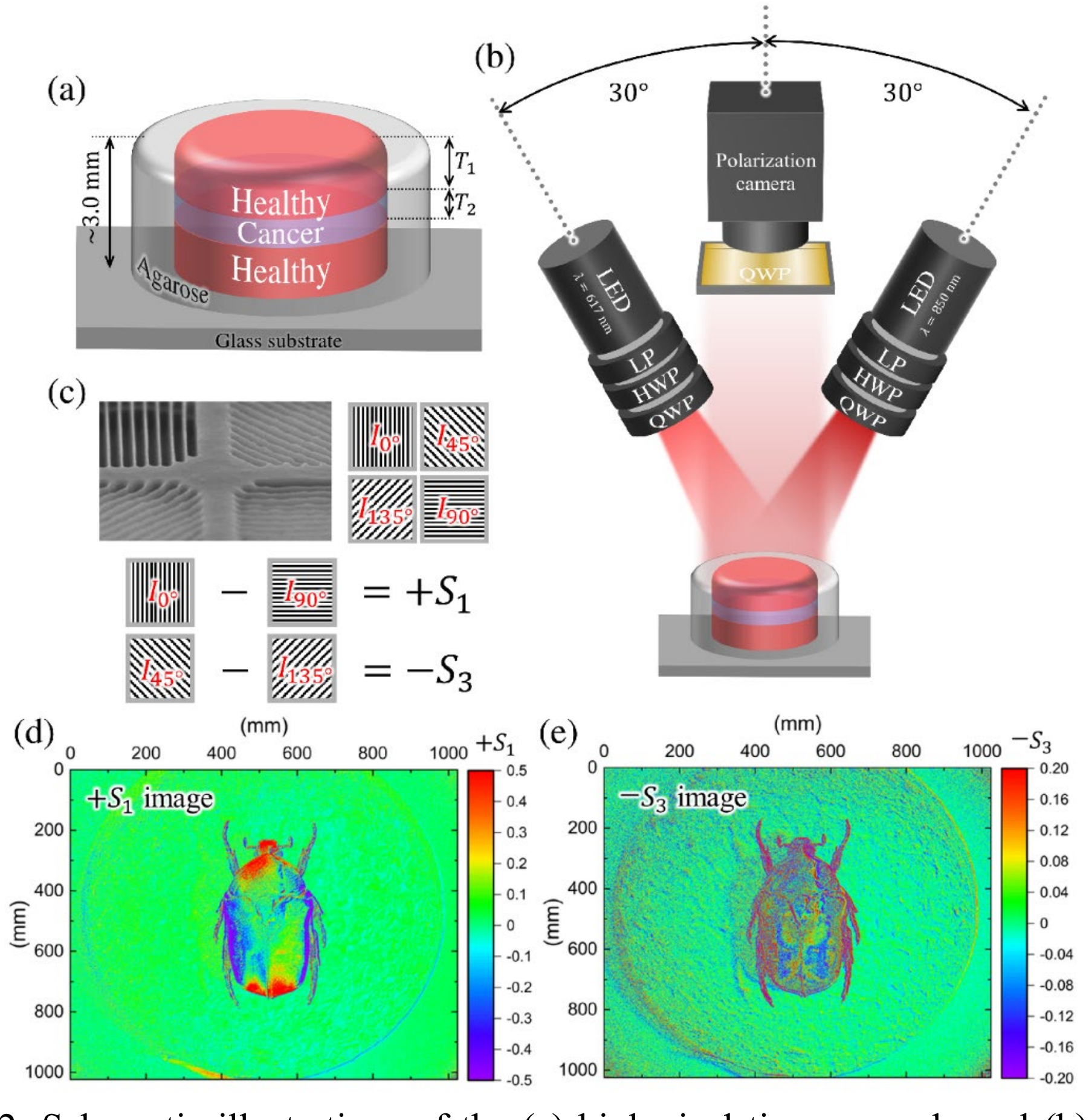

Figure 2: Schematic illustrations of the (a) biological tissue sample and (b) optical setup. (c) Microgrid patterns on the polarization camera and the relationships between the Stokes parameters and the polarized pixels. (d) $S_1$ and (e) $-S_3$ polarization images of a scarab under unpolarized light illuminations.

The polarization imaging system consisted of a polarization-sensitive camera (Toshiba Teli Corporation's BU505MZ-ES USB 3.0 Polarsens camera, a 2/3-inch Sony CMOS Pregius Polarsens sensor of model IMX250MZR Integrated with 4-Directional Wire Grid Polarizer Array [19]) with an 8MP lens (FUJINON HF3520-8M 1:2.0 / 35mm) and a QWP corresponding to the wavelengths of the irradiation CPL. The polarization imaging camera had $2 \times 2$ microgrid array patterns in which four neighboring pixels were polarizeed at $0°$, $45°$, $90°$, and $135°$, as depicted in Figure 2(c) [20]. The arrays of these polarized pixels enabled simultaneous measurement of the intensities passing through the polarizers with the specified directions, which were defined as $I_{0°}$, $I_{45°}$, $I_{90°}$, and $I_{135°}$. The relative intensities of the $S_1$ and $S_2$ components were obtained by taking the differences $I_{0°} - I_{90°}$ and $I_{45°} - I_{135°}$, respectively, yielding $S_1$ and $S_2$ images. Polarized light expressed by $\boldsymbol{S}$ can be converted into $\boldsymbol{S'}$ by a QWP according to the following equation:

$$\boldsymbol{S}' = \boldsymbol{M}_{QWP}(\theta) \cdot \boldsymbol{S}$$

$$\begin{pmatrix} S_0{}' \\ S_1{}' \\ S_2{}' \\ S_3{}' \end{pmatrix} = \begin{pmatrix} 1 & 0 & 0 & 0 \\ 0 & \cos^2 2\theta & \sin 2\theta \cos 2\theta & \sin 2\theta \\ 0 & \sin 2\theta \cos 2\theta & \sin^2 2\,\theta & -\cos 2\theta \\ 0 & -\sin 2\theta & \cos 2\theta & 0 \end{pmatrix} \begin{pmatrix} S_0 \\ S_1 \\ S_2 \\ S_3 \end{pmatrix}, \quad (1)$$

where $\theta$ is the angle formed by the fast axis of the QWP and a vertical line. Therefore, when viewed through a QWP with $\theta = 0°$, the $S_1$ and $S_2$ images are converted into $S_1$ and $-S_3$ images according to the following equation:

$$\boldsymbol{S}' = \boldsymbol{M}_{QWP}(\theta = 0°) \cdot \boldsymbol{S},$$

$$\begin{pmatrix} S_0{}' \\ S_1{}' \\ S_2{}' \\ S_3{}' \end{pmatrix} = \begin{pmatrix} +1 & 0 & 0 & 0 \\ 0 & +1 & 0 & 0 \\ 0 & 0 & 0 & -1 \\ 0 & 0 & +1 & 0 \end{pmatrix} \begin{pmatrix} S_0 \\ S_1 \\ S_2 \\ S_3 \end{pmatrix} = \begin{pmatrix} S_0 \\ +S_1 \\ -S_3 \\ S_2 \end{pmatrix}. \quad (2)$$

Therefore, we obtained $-S_3$ images using the circular polarization imaging system shown in Figure 2 (b). In this study, we took a combined snapshot, separated it into four polarized images, and obtained $S_1$ and $S_3$ images by subtraction between them in the same array.

For operational confirmation of CPL imaging, we took polarized images of a scarab, which exhibits selective reflection of left-handed CPL exclusively [21], under unpolarized light illumination. Figures 2 (d) and (e) show the $S_1$ and $-S_3$ images of the scarab. The $S_1$ values are distributed according to the curvature of the exocuticle. Meanwhile, the $-S_3$ values on scarab deviate from zero, indicating that one polarity of the CPL is detected more than the other, even though the unpolarized light contains both polarities of CPL equally. The DOCP values are calculated using $S_3/S_0$. However, because the $S_0$ component, which represents the total intensity, could not be accurately determined with the polarization camera, we defined $(I_{0°} + I_{45°} + I_{90°} + I_{135°})/2$ as the pseudo total intensity and used this pseudo intensity to calculate the DOCP in this study.

Therefore, in this study, DOCP values are defined as;

$$\mathrm{DOCP} = \frac{S_3}{S_0} \cong \frac{-2(I_{45^\circ} - I_{135^\circ})}{I_{0^\circ} + I_{45^\circ} + I_{90^\circ} + I_{135^\circ}}. \tag{3}$$

## III. Results and discussions

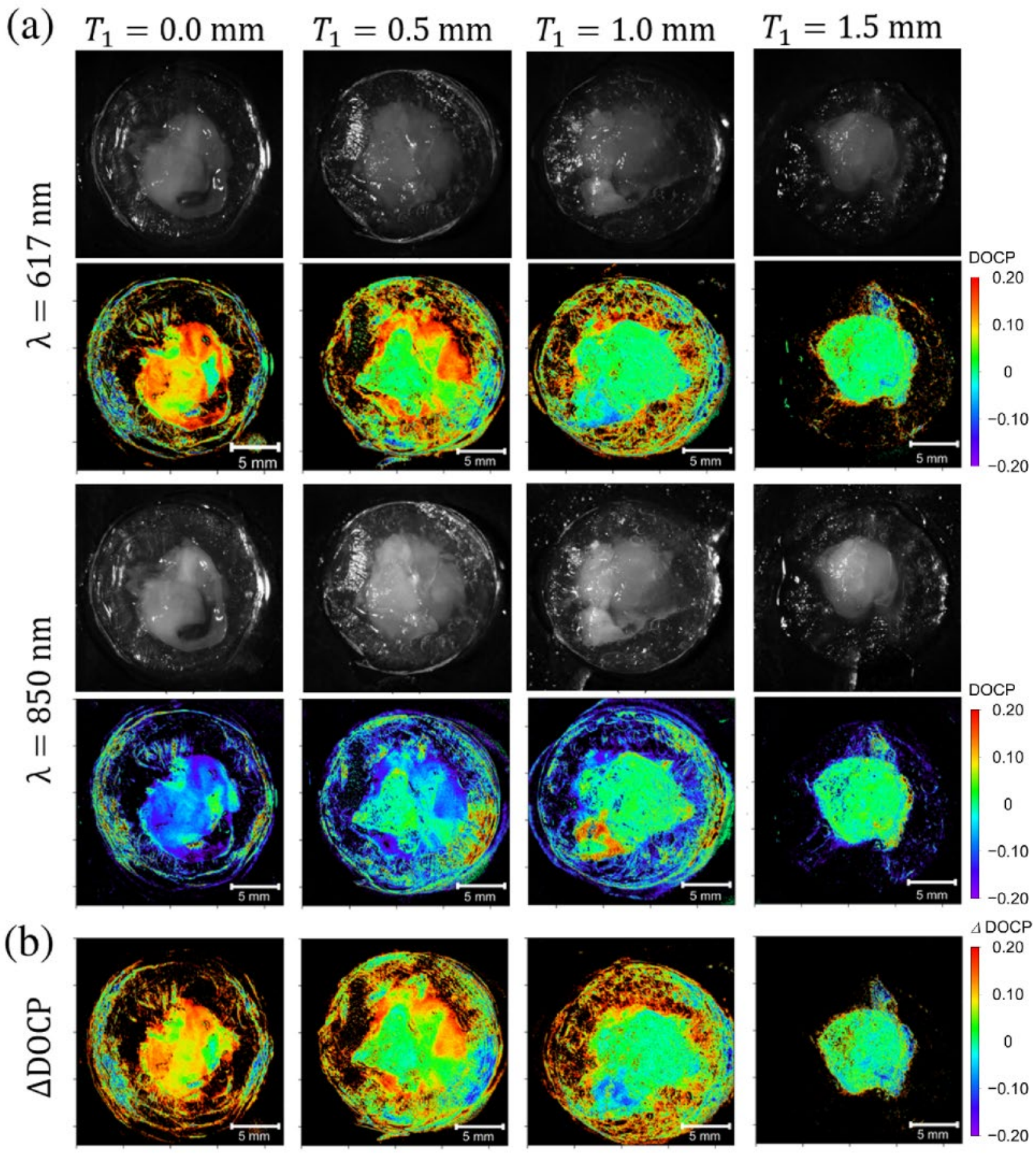

Figure 3: (a) Raw images and DOCP distribution images captured with the polarization imaging system for the samples with different $T_1$ and a fixed $T_2$ of 1.0 mm. The images captured with $\lambda = 617$ nm and 850 nm are in the upper and lower half-rows, respectively, and those for the samples with $T_1 = 0.0,\ 0.5,\ 1.0,$ and 1.5 mm are arranged sequentially from the left. (b) Images of the differences in the DOCP values (ΔDOCP) taken with the two wavelengths

Figure 3 shows the raw images and DOCP distribution images captured with the polarization imaging system for samples with different $T_1$ and a fixed $T_2$ of $1.0\ \mathrm{mm}$, which corresponds to samples 1, 4, 5, and 6, as shown in Table 1. The images captured at $\lambda = 617\ \mathrm{nm}$ and $850\ \mathrm{nm}$ are in the upper and lower half-rows, respectively, and those for the samples with $T_1 = 0.0,\ 0.5,\ 1.0$ and $1.5\ \mathrm{mm}$ are arranged sequentially from the left. In the raw monochromatic images, the whitish parts represent the biological tissues, and the surrounding transparent parts are agarose gels used to fix the tissues. The average DOCP values in the tissue area in each image are shown in Figure 4 (a) with red and blue closed squares for 617 and 850 nm, respectively. The error bars indicate the standard deviations among the same area of the tissue. When $T_1$ increases, the average DOCP values decrease and increase monotonically in the 617 nm and 850 nm cases, respectively. These values also include contributions from the surface reflection, surface roughness, and other factors. Figure 3 (b) shows the images of the differences in DOCP values taken with two wavelengths (ΔDOCP) to eliminate these unnecessary contributions, which are distribution images only with the contributions of CPL scattering. The ΔDOCP values are obtained by $\Delta\mathrm{DOCP} = \mathrm{DOCP}(617\ \mathrm{nm}) - \mathrm{DOCP}(850\ \mathrm{nm})$. Figure 4 (b) shows the cancer depth dependence of the ΔDOCP values obtained from Figure 3 (b). For comparison with simulations for similar configurations, Figure 4 (c) and (d) show the calculated DOCP values for the two wavelengths and ΔDOCP values extracted from the data shown in ref. [15]. The behaviors of the DOCP value have similar tendencies in increase and decrease but the variations in values are attributed to surface disturbances.

To assess the sensitivity with which the buried cancer layer was detected, the DOCP values were compared among the samples with different thickness of cancer, $T_2$ and a fixed $T_1$ of 0.5 mm, which correspond to samples 2, 3, and 4 shown in Table 1. Figure S1 in the supporting file provides the raw and DOCP distribution images, and Figure 5 (a) illustrates the ΔDOCP distribution images derived using the same methodology. The average DOCP and ΔDOCP values as functions of $T_2$ are shown in Figure 5 (b) and (c), respectively. Seemingly, the obtained data depend on the volume of the cancerous layer; however, the variations in the $T_2$ dependence are smaller by approximately one digit than those in the $T_1$ dependence. Because the penetration depths of light with $\lambda = 617$ nm and 850 nm are almost the same (approximately 2.5 mm), the cancer layer in these samples is included in the scattering volume of the irradiated CPL. However, as the depth increases, the amount of light that reaches it decreases. Accordingly, the sensitivity for detecting the upper boundary between the upper healthy layer and the lower cancerous layer is higher than that for detecting the lower boundary of the cancerous layer with the bottom layer.

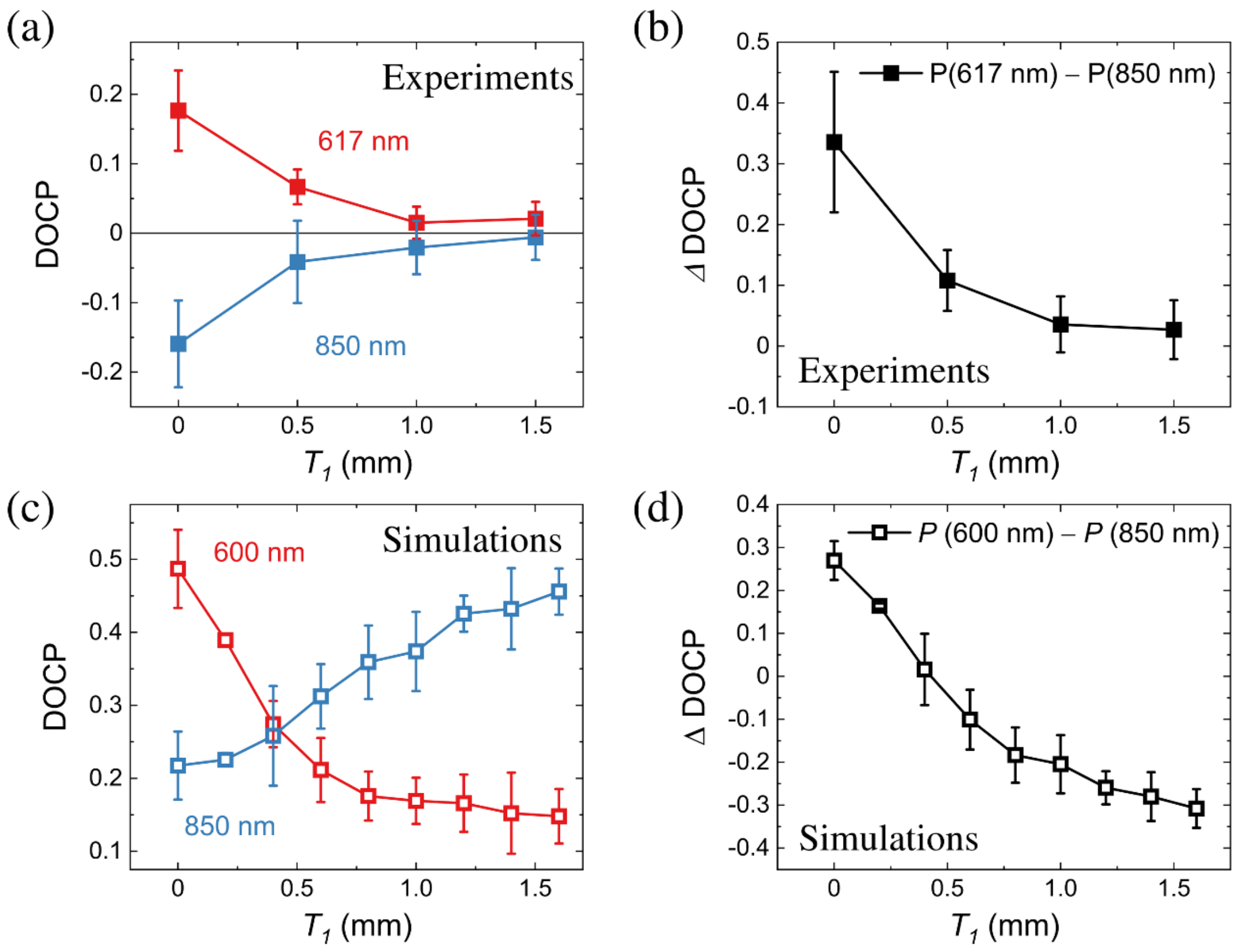

Figure 4: (a) $T_1$ dependence of the average DOCP values in the tissue area in the DOCP distribution images shown in Figure 3 with $\lambda =$ (red) 617 nm and (blue) 850 nm. (b) Differences in the DOCP values (ΔDOCP) between the two wavelengths. The calculation results for the (c) DOCP and (d) ΔDOCP values in the similar configurations, which are extracted from the data shown in [15].

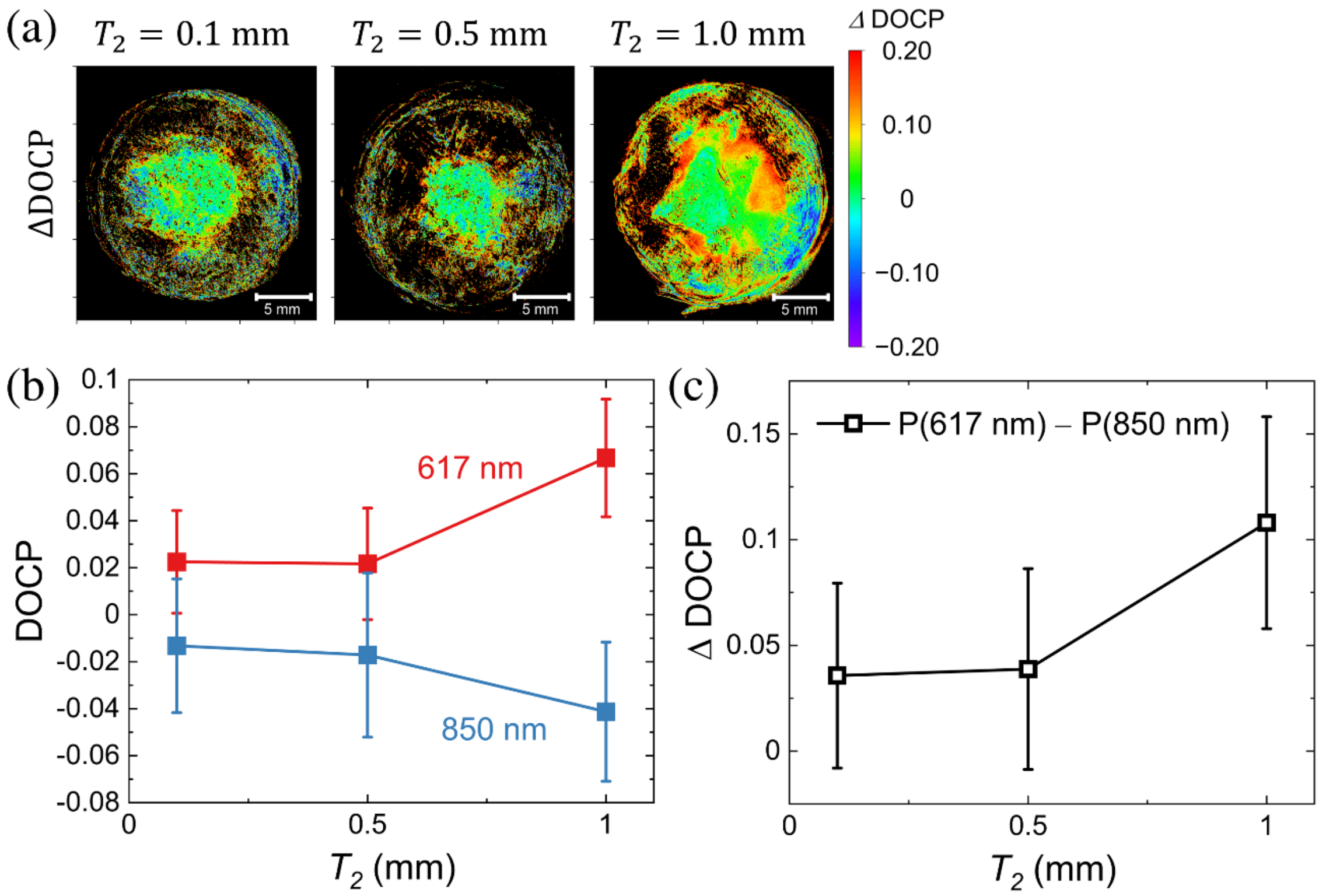

Figure 5: (a) Images of the differences in the DOCP values between the two wavelengths. (b) $T_2$ dependence of the average DOCP value in the tissue area in the DOCP distribution images shown in Figure S1 with $\lambda =$ (red) 617 nm and (blue) 850 nm. (b) Differences in the DOCP values (ΔDOCP) between the two wavelengths.

## IV. Conclusions

We experimentally identified cancer tissues buried beneath healthy tissues using

CiPLS. Unexposed cancer tissues consisting of healthy/cancerous/healthy layers were artificially prepared with various thicknesses of the top healthy layer and the cancerous layer. CPL with $\lambda = 617$ nm and $850$ nm was irradiated onto the entire tissue, and the distributions of the circular polarization of the scattered light was visualized through a QWP with a polarization imaging camera. The thickness variation of the top layer, specifically the depth variation of the cancerous layer, caused significant changes in the DOCP values. Notably, the directions of these changes were opposite for different wavelengths: a decrease for 617 nm irradiation and an increase for 850 nm irradiation with increasing depth. The differences in the DOCP values between these wavelengths showed a monotonic decrease with the depth of the cancer layer. On the other hand, the volume of the cancer layer at a depth of 1.0 mm contributed little to the DOCP values when the thickness of the cancer layer was at least 0.1 mm. In conclusion, the depth of unexposed cancer lying within the optical penetration depth can be evaluated using a circular polarization imaging system based on the CiPLS method. The thickness of the squamous epithelium, which varies from 0.3 to 0.7 mm in CIN1 and CIN2 [22], is within the optical penetration depth.

Artificially layered tissue samples were used in this study as the first step towards non-invasive optical diagnosis of early-stage cervical cancer. The next step will be experimental demonstration with actual cervical tissues having various low-grade CIN. Moreover, the correlations between the obtained DOCP values and the pathological properties of actual cervical tissues, such as density, anisotropy, and structure of cell nuclei, need to be examined in detail. Concurrently, computational analyses that include more realistic parameters are required to enhance the accuracy of the calculations. For example, the cell density and optical parameters vary according to the depth, as well as the detailed structure of the tissue, including the basal layer and stroma. Improvements in optical devices for CPL irradiation and detection (imaging) are also required. The current irradiation system, which is composed of LEDs, wave plates, and lenses, offers CPL irradiation with $30°$ of the incident angle. Owing to variations in the intensity and circular polarization along the radial direction, the reliability of the system is compromised. One of the solutions for this problem is to use metalenses designed with uniform polarization distributions, as well as intensity distribution. Moreover, the integration of a circular polarization imaging system on the tip of an endoscopic apparatus requires miniaturization and simplification to enhance the functionality and ease of use of the device in medical procedures. In the current imaging system, the $S_1$ images are underutilized; however, they yield information about the inclination of the tissue surface in the same manner as tilt ellipsometry [23]. Angular information would improve the sensing accuracy of the depth measurements. After overcoming these challenges, we aim to develop a fine and flexible colposcope capable of CiPLS imaging. This medical apparatus will provide a method of diagnosing the cervix that is less burdensome on patients, which will enhance the effectiveness of periodic inspections for the early detection of cervical cancer. This research can pave the way for more accurate and non-invasive diagnostic techniques in the future.

**Ethical approval**

Ethical approval was not required to carry out this study.

**Disclosures**

The authors declare that there is no conflict of interest related to this article.

**Acknowledgments**

This study was partially supported by KAKENHI (Nos. 19H04441, 22H03921, 23K25175, and 25K03438) of the Japan Society for the Promotion of Science (JSPS), Cooperative Research Project of the Research Center for Biomedical Engineering, Uehara Memorial Foundation, and grants from Kitasato university (No. 1768 in 2022–2025). The authors acknowledge Mr. Mitsuo Fukuno and Akane Saito for fruitful discussion and technical support at Kitasato University, Japan.

**Code and Data Availability**

The data and code supporting the findings reported in this article are publicly available and can be obtained from the authors upon reasonable request from nishizawa.nozomi@kitasato-u.ac.jp.

## Reference

[1] L. Bruni, G. Albero, B. Serrano, M. Mena, J. J. Collado, J. Gómez, F. X. Bosch, S. de Sanjose. ICO/IARC Inf Cent HPV Cancer (HPV Inf Centre). *Human papillomavirus and related diseases report in world.* Summary Report 10 March 2023. Available from: www.hpvcentre.com.

[2] M. H. Stoler, M. Schiffman, JAMA **285**, 1500-1505 (2001).

[3] F. Bray, A. Jemal, N. Grey, J. Ferlay, D. Forman, Lancet Oncol. **13**, 790 (2012).

[4] A. B. Sravani, V. Ghate, S. Lewis S, Biol. Trace Elem. Res. **201**, 1026-1050 (2023).

[5] P. Guo, H. Almubarak, K. Banerjee, R. J. Stanley, R. Long, S. Antani, G. Thorma, Ro. Zuna, S. R. Franzier, R. H. Moss, W. V. Stoecker, J. Phothol. Inform. 7, 51 (2016).

[6] P. Nieminen, M. M. Lallio, M. Hakama, Obstetrics and gynecology **85**, 1017 (1995).

[7] A. Ullal, M. Roverts, J. N. Bulmer, M. E. Mathers, V. Wadehra, Cytopathology **20**, 359 (2008).

[8] N. Nishizawa, A. Hamada, K. Takahashi, T. Kuchimaru, H. Munekata, Jpn. J. Appl. Phys. **59**, SEEG03 (2020).

[9] N. Nishizawa, A. Esumi, Y. Ganko, J. Biomed. Opt. **29**, 07501 (2024).

[10] D. Bicout, C. Brosseau, A. S. Martinez, J. M. Schmitt, Phys. Rev. E **49**, 1767 (1994).

[11] F. C. MacKintosh, J. X. Zhu, D. J. Pine, D. A. Weitz, Phys. Rev. B **40**, 9342 (1989).

[12] C. F. Bohren and D. R. Huffman, *Absorption and Scattering of Light by Small Particles*, Wiley, New York (1983).

[13] B. Kunnen, C. Macdonald, A. Doronin, S. Jacques, M. Eccles, and I. Meglinski, J. Biophotonics **8**, 317 (2015).

[14] N. Nishizawa, B. Al-Qadi, T. Kuchimaru, J. Biophotonics **14**, e202000380 (2021).

[15] N. Nishizawa, T. Kuchimaru, J. Biophotonics **15**, e202200062 (2022).

[16] N. Nishizawa, Sl. Kawashima, B. Al-Qadi, T. Kuchimaru, H. Munekata, Proc. SPIE **11521**, 1152114 (2020).

[17] M. R. Maskey, M. Fukuno, A. Saito, A. Exumi, T. Kuchimaru and N. Nishizawa, Proc. SPIE *in press*.

[18] E. Collet, *Field Guide to Polarization*, SPIE, Bellingham, WA (2005).

[19] Sony Semiconductor Solutions Corporation, "Polarization image sensor with four-directional on-chip polarizer and global shutter function," 2021, https://www.sony-semicon.co.jp/e/products/IS/industry/product/polarization.html.

[20] C. Lane, D. Rode, T. Rösgen, Appl. Opt. **60**, 8435 (2021).

[21] L. T. McDonald, E. D. Finlayson, B. D. Wilts, P. Vukusic, Interface focus 7, 20160129 (2017).

[22] I. Ghosh, S. Mittal, D. Banerjee, N. Chowdhury, P. Basu, Int. J. Gynecol. Pathol. **35**, 269-274 (2015).

[23] T. Tsuru, Opt. Exp. **21**(5), 6625 (2013).

## Figure Caption

Figure 1:

Schematic illustrations of the (a) cervix and uterus with the tissue structure near the SCJ and (b) cervical pre-cancer and cancer progression: normal, CIN1, CIN2, CIN3, and invasive cancer. Representative hematoxylin and eosin images of the squamous epithelium in (c) normal (d) CIN1, (e) CIN2, and (f) CIN3 cases.

Figure 2:

Schematic illustrations of the (a) biological tissue sample and (b) optical setup. (c) Microgrid patterns on the polarization camera and the relationships between the Stokes parameters and the polarized pixels. (d) $S_1$ and (e) $-S_3$ polarization images of a scarab under unpolarized light illuminations.

Figure 3:

(a) Raw images and DOCP distribution images captured with the polarization imaging system for the samples with different $T_1$ and a fixed $T_2$ of $1.0\,\mathrm{mm}$. The images captured with $\lambda = 617\,\mathrm{nm}$ and $850\,\mathrm{nm}$ are in the upper and lower half-rows, respectively, and those for the samples with $T_1 = 0.0,\ 0.5,\ 1.0,$ and $1.5\,\mathrm{mm}$ are arranged sequentially from the left. (b) Images of the differences in the DOCP values (ΔDOCP) taken with the two wavelengths

Figure 4:

(a) $T_1$ dependence of the average DOCP values in the tissue area in the DOCP distribution images shown in Figure 3 with $\lambda =$ (red) $617\,\mathrm{nm}$ and (blue) $850\,\mathrm{nm}$. (b)

Differences in the DOCP values (ΔDOCP) between the two wavelengths. The calculation results for the (c) DOCP and (d) ΔDOCP values in the similar configurations, which are extracted from the data shown in [15].

Figure 5:

(a) Images of the differences in the DOCP values between the two wavelengths. (b) $T_2$ dependence of the average DOCP value in the tissue area in the DOCP distribution images shown in Figure S1 with $\lambda =$ (red) $617\ \mathrm{nm}$ and (blue) $850\ \mathrm{nm}$. (b) Differences in the DOCP values (ΔDOCP) between the two wavelengths.